\documentclass[11pt,a4paper]{article}
\usepackage[margin=1in]{geometry}
\usepackage{amsmath,amsthm,amssymb}
\usepackage{graphicx}
\usepackage{dcolumn}
\usepackage{bm}
\usepackage{color}
\usepackage{booktabs}
\usepackage[numbers,sort&compress]{natbib}
\usepackage[pdftitle={Measurement-Free Toric-Code Memory in Array Globally Controlled Rydberg Array},pdfauthor={Han Wang, Yusheng Zhao, Xiu-Hao Deng, Jinguo Liu}]{hyperref}
\usepackage{cleveref}
\usepackage{subfigure}
\usepackage{braket}
\usepackage{placeins}
\usepackage{enumitem}
\usepackage{authblk}

\IfFileExists{orcidlink.sty}{\usepackage{orcidlink}}{%
}

\newenvironment{acknowledgments}{\section*{Acknowledgements}}{}

\newcommand{\avg}[1]{\langle #1 \rangle}

\title{Self-Correcting Toric-Code Memory in a Globally Controlled Rydberg Array}

\author[1]{Han Wang}
\author[1]{Yusheng Zhao}
\author[2]{Xiu-Hao Deng\thanks{Corresponding author: \texttt{dengxiuhao@iqasz.cn}}}
\author[1]{Jinguo Liu\thanks{Corresponding author: \texttt{jinguoliu@hkust-gz.edu.cn}}}
\affil[1]{Advanced Materials Thrust, Function Hub, The Hong Kong University of Science and Technology (Guangzhou), Guangdong 511400, China}
\affil[2]{International Quantum Academy, and Shenzhen Branch, Hefei National Laboratory, Shenzhen,518048, China.}

\date{\today}

\begin{document}

\maketitle

\begin{abstract}
% {Abstract is $\sim$196 words; npj QI Article limit is 150 words --- trim before submission.}
The central prerequisite of any fault-tolerant quantum architecture is a quantum memory: a block of encoded physical qubits whose logical state is actively preserved against noise across many rounds of error correction. In neutral-atom Rydberg arrays, realizing such a memory is obstructed not by the entangling gates themselves, which are already fast and high-fidelity, but by the auxiliary operations that a conventional error-correction cycle requires — mid-circuit fluorescence measurement, inter-zone atom transport, and locally focused single-qubit addressing. Each of these introduces latency, atom loss, or optical crosstalk that exceeds the cost of the underlying gates by orders of magnitude. These costs accumulate cycle after cycle, progressively degrading the very logical information the code is meant to protect. Here we propose a protocol that stabilizes a toric-code quantum memory without moving, measuring or local addressing atoms. The key is to use a three-species Rydberg atom array for the complete stabilizer cycle, including syndrome extraction, coherent correction, and ancilla reset, under global, species-selective laser pulses. Numerical simulation of a $4\!\times\!4$ rotated toric code shows a longer qubit lifetime when the physical error rate is below a pseudo-threshold $p^\star\!\approx\!0.034$. The scheme offers a concrete, hardware-efficient route to topological quantum memory in neutral-atom platforms.
\end{abstract}

%%%%%%%%%%%%%%%%%%%%%%%%%%%%%%%%%%%%%%%%%%%%%%%%%%%%%%%%%%%%%%%%%%%%%%%%%%%%%%%
\section{Introduction}
\label{sec:introduction}
%%%%%%%%%%%%%%%%%%%%%%%%%%%%%%%%%%%%%%%%%%%%%%%%%%%%%%%%%%%%%%%%%%%%%%%%%%%%%%%

Neutral-atom arrays manipulated by optical tweezers and Rydberg interactions~\cite{saffman2010quantum,browaeys2020many,kaufman2021quantum,morgado2021quantum} now combine fast gates with long coherence. Entangling gates have reached the $\sim\!100$--$300$~ns timescale with fidelities exceeding $99.5\%$~\cite{evered2023high,ma2023high}, and most recently $\sim\!99.941\%$~\cite{evered2026nonlocal}. Ground-state hyperfine coherence extends to seconds~\cite{barnes2022assembly,jenkins2022ytterbium}, and reconfigurable arrays of hundreds to thousands of atoms are now routine~\cite{ebadi2021quantum,scholl2021quantum,bluvstein2024logical}. These fidelities correspond to per-gate error rates as low as $\sim\!6\times10^{-4}$, below the $\sim\!1\%$ fault-tolerance threshold of the surface and toric codes~\cite{fowler2012surface}. Yet a complete error-correction cycle requires far more than entangling gates: it demands mid-circuit measurement, atom transport, and locally addressed single-qubit operations. These auxiliary operations are primary sources of computing time and qubit error. Fluorescence-based mid-circuit readout requires milliseconds~\cite{norcia2023midcircuit,graham2023midcircuit}, during which unmeasured qubits continue to decohere and scattered photons induce crosstalk~\cite{martinez2016real,cong2022hardware}. Shuttling atoms between memory and measurement zones~\cite{bluvstein2022quantum,bluvstein2024logical} adds tens to hundreds of microseconds of latency and heats the motional state, introducing motional errors. Local single-qubit addressing via tightly focused beams introduces optical crosstalk and a hardware overhead that scales with system size~\cite{weitenberg2011single,xia2015randomized}. 

To reduce overhead of auxiliary operations, several schemes have been proposed: In order to remove mid-circuit measurement, Heussen \emph{et al.}~\cite{heussen2024measurementfree} replace the classical decoder with a coherent quantum circuit, and Petiziol \emph{et al.}~\cite{petiziol2024floquet} Floquet-engineer the toric-code Hamiltonian on superconducting qubits to encode its ground state naturally. But both of them still rely on local addressing or atom movement. In order to 
avoid atom shuttling, Singh \emph{et al.}~\cite{Singh2023} exploits a dual-species Rb/Cs array so that mid-circuit readout of Cs spectator qubits does not disturb the Rb data atoms. However, the projective ancilla measurement is still required. We instead ask whether mid-circuit measurement, atom transport, and local addressing can all be removed simultaneously.

Multi-species neutral-atom arrays are now an experimentally mature platform. 
Element-selective optical tweezers support dual-element Rb--Cs registers of up to 512 sites with negligible interspecies crosstalk~\cite{singh2022dual}; the same wavelength-selective trapping has been extended to Rb--K~\cite{wei2024dual,angonga2022gray}, and simultaneous magneto-optical trapping of three atomic species was demonstrated nearly two decades ago~\cite{taglieber2006three}. 
Together with theoretical F\"orster catalogues for the relevant alkali pairs~\cite{PhysRevA.92.042710,otto2020interspecies,ireland2024interspecies} and metastable-manifold mid-circuit control~\cite{lis2023midcircuitoperationsusingomgarchitecture,ma2023high}, these advances make species selectivity a practical substitute for local optical addressing and the experimental basis for our three-species protocol.

Based on these advances, we propose a protocol for stabilizing a toric-code quantum memory with atom-species selective global pulses; no atom movement, local addressing or mid-circuit measurements.
We validate the protocol at the gate, cycle, and code levels, finding a pseudo-threshold of $p^\star\!\approx 0.034$ for the full $4\!\times\!4$ rotated toric code under depolarizing noise.
The key challenge is global pulse design: the induced unitaries cannot be easily decomposed into local ones, even when only nearest-neighbor Rydberg blockade interactions are considered.
Traditional gate level pulse design techniques do not trivially apply here.
Designing a proper pulse sequence to achieve the wanted stabilizer Hamiltonian is a key contribution of our work.

The remainder of the paper is organized as follows. \Cref{sec:system} introduces the toric-code memory and the three-species global-pulse protocol that implements one correction cycle. \Cref{sec:results} benchmarks the per-triple pulse unitaries and simulates the full $4\!\times\!4$ rotated toric-code cycle under a per-round data-qubit depolarizing noise model. \Cref{sec:discussion} discusses scope, limitations, and outlook. 

%%%%%%%%%%%%%%%%%%%%%%%%%%%%%%%%%%%%%%%%%%%%%%%%%%%%%%%%%%%%%%%%%%%%%%%%%%%%%%%
\section{Results}
\label{sec:results-main}
%%%%%%%%%%%%%%%%%%%%%%%%%%%%%%%%%%%%%%%%%%%%%%%%%%%%%%%%%%%%%%%%%%%%%%%%%%%%%%%

\subsection{System and protocol}
\label{sec:system}
\label{sec:scheme-overview}% retained so the Methods cross-ref still resolves

\begin{figure*}[htbp]
    \centering
    \includegraphics[width=0.8\linewidth]{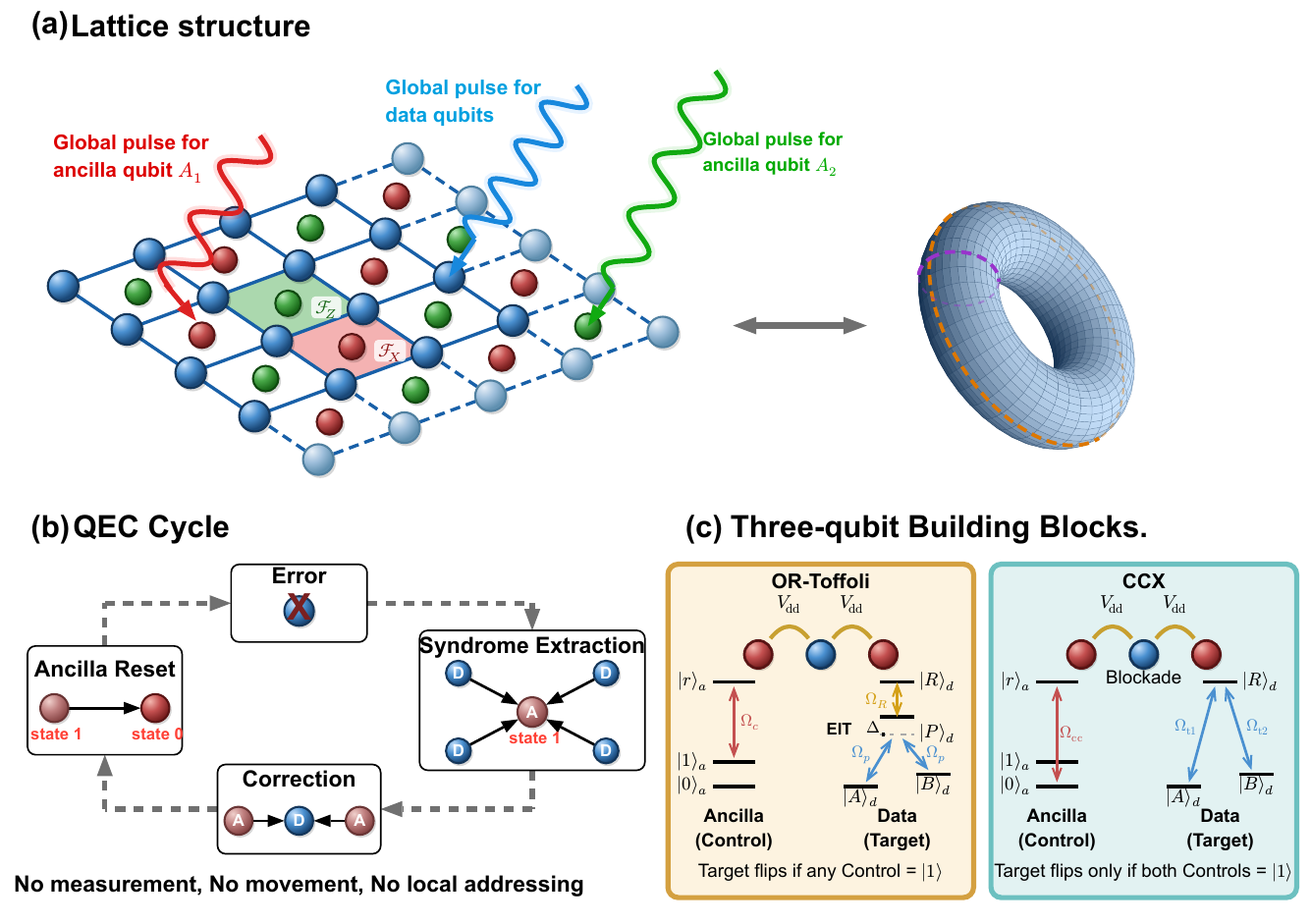}
    \caption{Overview of the measurement-free, movement-free quantum-memory protocol on a multi-species Rydberg atom array. (a) Lattice: blue spheres, data qubits (species~D) at lattice vertices; red spheres, ancilla species A$_1$ at $Z$-plaquette centers; green spheres, ancilla species A$_2$ at $X$-plaquette centers. The green and red shaded region represents the different plaquette regions $\mathcal{F}_X$ and $\mathcal{F}_Z$. Three species-selective global laser fields address D, A$_1$, and A$_2$ independently; the torus sketch on the right emphasizes the periodic-boundary layout, with the dashed magenta and orange loops marking the two independent non-contractible cycles that support the logical operators $\bar X_{1,2}$ and $\bar Z_{1,2}$. (b) Logical cycle: errors on the data qubits are coherently mapped onto two adjacent ancillas, a conditional correction gate heals the data, and a global optical-pumping pulse resets the ancillas to $\ket{0}$---all without measurement, atom movement, or local addressing. (c) Gate building blocks on a single (D, A, A) triple --- data atom $d$ as target, the two ancillas $a_1, a_2$ as controls. Left: the three-qubit \textsc{or-toffoli} gate (also called the dual Toffoli~\cite{moraga2020dualtoffoli,moraga2021ortoffoli}) flips the target whenever \emph{at least one} control is in $\ket{1}$, i.e.\ it implements the Boolean-exponent Pauli $X_d^{a_1 \vee a_2}$ (an $X$ on data atom $d$ that fires iff $a_1 \vee a_2 = 1$); we realize it by electromagnetically-induced transparency (EIT) on the target combined with Rydberg blockade from the controls. Right: the three-qubit \textsc{ccx} (standard Toffoli) flips the target only when \emph{both} controls are in $\ket{1}$, i.e.\ $X_d^{a_1 \wedge a_2}$; we realize it by direct resonant Rydberg coupling.}
    \label{fig:model}
\end{figure*}

The protocol stabilizes Kitaev's toric code~\cite{kitaev2003fault,dennis2002topological,fowler2012surface} on an $L\!\times\!L$ square lattice rotated $45^\circ$ with periodic boundary conditions (\Cref{fig:model}(a)). Data qubits (species~$D$) sit on lattice vertices (blue spheres); four neighboring data qubits form a plaquette (shaded areas). Alternating plaquettes host two ancilla species $A_1$ and $A_2$, forming the checkerboard sublattices $\mathcal{F}_X$ (red shaded area) and $\mathcal{F}_Z$ (green shaded area) that carry the two stabilizer families
\begin{equation}
    A_s = \prod_{i \in \partial s} X_i \;\; (s \in \mathcal{F}_X), \qquad B_p = \prod_{i \in \partial p} Z_i \;\; (p \in \mathcal{F}_Z),
    \label{eq:stabilizers}
\end{equation}
where $\partial s$ (resp.\ $\partial p$) denotes the four data qubits at the vertices of plaquette $s$ (resp.\ $p$). Because the code is a Calderbank--Shor--Steane (CSS) code, each data qubit belongs to exactly two $A_s$ and two $B_p$ plaquettes, so $X$ and $Z$ errors can be detected in two independent \emph{sub-cycles}, each using a single stabilizer type. The encoding hosts $k=2$ logical qubits with code distance $d_{\rm code}=L$ set by the shortest non-contractible loop on the torus (shown in \Cref{fig:model}(a)). This CSS splitting is the structural reason our protocol carries \emph{two} ancilla species (one per sub-cycle) in addition to the data species with species-selective global pulse. Each sub-cycle is then a coherent three-step operation built from these global pulses alone (\Cref{fig:model}(b)): (i) \emph{syndrome extraction}, mapping the stabilizer parities onto the ancillas (initially in $\ket{0}$); (ii) \emph{conditional correction}, applying a Pauli flip to the data conditioned on the ancilla state; and (iii) \emph{ancilla reset}, returning all ancillas to $\ket{0}$ via an optical-pumping pulse before the next cycle. The remainder of this section builds these steps: \Cref{sec:syndrome_design} designs the syndrome-extraction pulses on the three-atom triple; \Cref{sec:correction_reset} closes the cycle with the correction sweep and species-selective ancilla reset.

\subsubsection{Syndrome extraction pulse design}
\label{sec:syndrome_design}
\begin{figure*}[!htbp]
    \centering
    \includegraphics[width=0.85\linewidth]{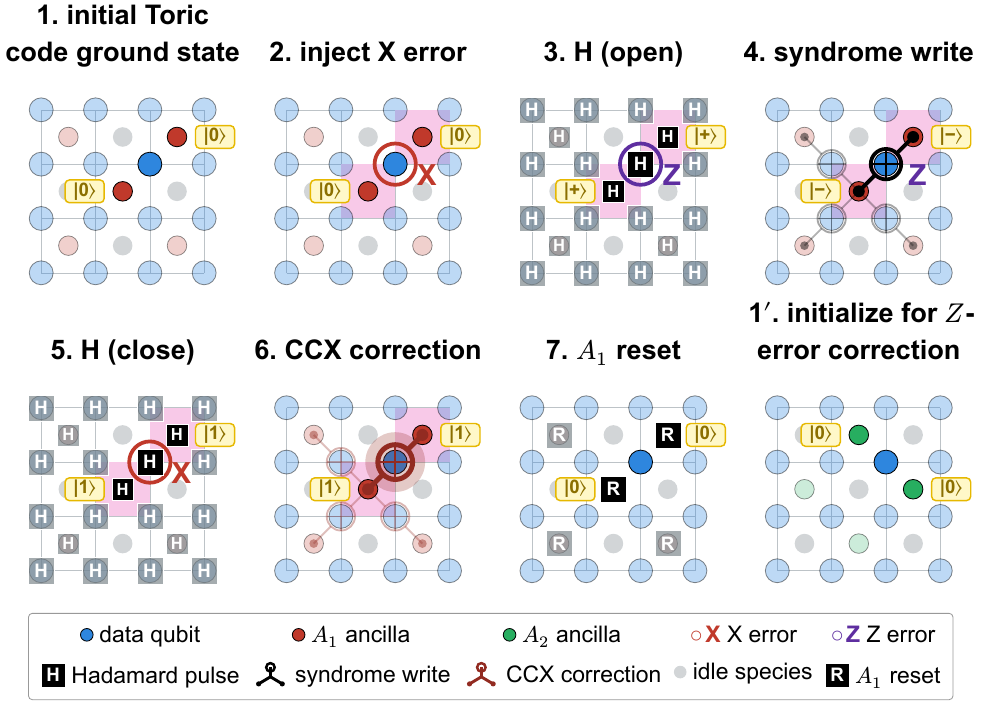}
    \caption{Tick-by-tick walkthrough of one $X$-error correction sub-cycle on the $4\!\times\!4$ rotated toric code, showing how a single injected $X$-error propagates through the measurement-free protocol. Blue discs represent data qubits and red/green discs represent A$_1$/A$_2$ ancillas (faint dots are not affected by injected error).
    \textbf{1.}~Toric-code ground state $\ket{\psi_0}$, all stabilizers $+1$, ancillas in $\ket{0}$.
    \textbf{2.}~An $X$-error is injected on data qubit $D^\star$ (red circle, label \textbf{X}); the two adjacent $B_p$ plaquettes (pink shading) now carry $-1$ eigenvalues, but no syndrome has yet been written to the ancillas.
    \textbf{3.~H~(open)}: a global Hadamard pulse rotates all A$_1$ ancillas to $\ket{+}$ and all data qubits by $H$; under this rotation the injected error on $D^\star$ transforms $X\!\to\!Z$ (purple label).
    \textbf{4.~Syndrome write}: a global \textsc{or-toffoli} sweep followed by a global \textsc{ccx} sweep maps the plaquette parity onto the two neighbor ancillas, taking them $\ket{+}\!\to\!\ket{-}$; the black star-on-cross icons mark the per-triple gate footprints (see legend).
    \textbf{5.~H~(close)}: a second global Hadamard closes the mapping, rotating ancillas $\ket{-}\!\to\!\ket{1}$: the syndrome is now physically present in the computational basis ($D^\star$ back $Z\!\to\!X$).
    \textbf{6.~CCX correction}: a global \textsc{ccx} pulse on every triple. The two adjacent ancillas in $\ket{1}$ act as joint controls and flip $D^\star$ back, healing the $X$-error while leaving the ancillas unchanged.
    \textbf{7.~$A_1$ reset}: a global optical-pumping pulse resets all A$_1$ ancillas to $\ket{0}$ (\textbf{R} markers), and the lattice is ready for the next sub-cycle.
    \textbf{1$'$.~Initialize for $Z$-error correction}: the symmetric next sub-cycle begins, now with A$_2$ (green) playing the role A$_1$ played above and a new pair of haloed ancillas; the remaining gates and Hadamard pattern are adapted to the $Z$ basis (not shown).}
    \label{fig:tick_walkthrough}
\end{figure*}
For the syndrome extraction process shown in \Cref{fig:model}(b), standard measurement-based implementation~\cite{fowler2012surface,acharya2023suppressing} uses a circuit of four CNOT gates coupling the ancilla to its nearest-neighbor data qubits via local addressing, mapping the data errors onto the ancilla.
In the lattice structure shown in \Cref{fig:model}(a), the sequential CNOT gates with local addressing can be expressed as:
\begin{equation}
\begin{split}
    |\mathbf{a}; \mathbf{d}\rangle \;&\mapsto\;
        \prod_{(i,j) \in S} X^{a_i}_{\mathrm{mid}(i, j)}
                            X^{a_j}_{\mathrm{mid}(i, j)}
        |\mathbf{a}; \mathbf{d}\rangle\\
    &=\;
        \prod_{(i,j) \in S} X^{a_i \oplus a_j}_{\mathrm{mid}(i, j)}
        |\mathbf{a}; \mathbf{d}\rangle,
\end{split}
\label{eq:CNOT--XOR}
\end{equation}
where $\mathbf{a}=(a_i)$ and $\mathbf{d}=(d_j)$ collect the ancilla and data computational-basis bits respectively, $S$ is the set of nearest-neighbor ancilla pairs sharing a data atom at the midpoint $\mathrm{mid}(i,j)$, and $X^{f(a_i,a_j)}$ denotes a controlled-$X$ that fires only when the Boolean function $f$ evaluates to true. 
This produces an $X$-parity for each plaquette. We then apply a Hadamard rotation that inverts the assignment: data become controls, ancillas become targets~\cite{acharya2023suppressing}.

However, local addressing is not permitted in our species-selective global pulse scheme. Therefore, the unit of pulse design is no longer chosen by a circuit compiler but forced by the lattice geometry. 
On the lattice of \Cref{fig:model}(a), each data qubit sits at the midpoint between \emph{two} ancillas of the active sublattice and is blockaded by both of them simultaneously. The natural unit is therefore the three-atom (data, ancilla, ancilla) triple (\Cref{fig:tick_walkthrough}), on which the global drive acts as a single coherent unitary.
Within each triple, each ancilla blockades the data atom without disturbing its partner; across triples, species-selective addressing and the negligible same-species ancilla--ancilla coupling keep cross-triple effects small (Methods~\ref{app:interactions}). The XOR parity achieved by the sequential CNOT operations in local addressing protocols~\cite{acharya2023suppressing} can therefore be expressed as the global operation acting on a three-atom unit:

\begin{equation}
\begin{split}
    |\mathbf{a}; \mathbf{d}\rangle \;&\mapsto\;
        \prod_{(i,j) \in S} X^{a_i \oplus a_j}_{\mathrm{mid}(i, j)}
        |\mathbf{a}; \mathbf{d}\rangle\\
    &=\;
        \prod_{(i,j) \in S} X^{a_i \lor a_j}_{\mathrm{mid}(i, j)}
        \prod_{(i,j) \in S} X^{a_i \land a_j}_{\mathrm{mid}(i, j)}
        |\mathbf{a}; \mathbf{d}\rangle,
\end{split}
\label{eq:xor_factorization}
\end{equation}
\Cref{eq:xor_factorization} is the Boolean identity $a\oplus b = (a\vee b)\oplus(a\wedge b)$ written for every $(i,j)\in S$. As operators on the three-atom Hilbert space, both factors are diagonal in the ancilla computational basis and act as $I$ or $X$ on the common data target; they therefore commute, and either firing order yields the same XOR parity. We name the two global pulses as follows. The first factor, $\prod_{(i,j)\in S} X^{a_i\vee a_j}_{\mathrm{mid}(i,j)}$, is implemented by a single global pulse we call the \emph{OR-Toffoli sweep} (\textsc{or-toffoli})~\cite{moraga2021ortoffoli} illustrated in \Cref{fig:model}(c,left) with pulse-level details in \Cref{app:pulse_details}. It flips the midpoint data atom whenever at least one of the two adjacent ancillas is in $\ket{1}$. 

The second factor, $\prod_{(i,j)\in S} X^{a_i\wedge a_j}_{\mathrm{mid}(i,j)}$, is implemented by the standard Toffoli on the same triple, which we call the \emph{CCX sweep} (\textsc{ccx}) [\Cref{fig:model}(c,right)]: it flips the midpoint data atom only when both adjacent ancillas are in $\ket{1}$. The composition of one OR-Toffoli sweep followed by one CCX sweep reproduces the XOR parity that sequential CNOTs would write. 
% The remainder of this subsection constructs each gate physically; 
The whole process is shown in \Cref{fig:tick_walkthrough}, Tick~4.

Hadamard rotations exchange control and target~\cite{PhysRevA.96.052320, acharya2023suppressing}. The $X$-sub-cycle, which detects $X$-errors via the $Z$-stabilizer, applies Hadamards to all data and ancilla qubits before and after the entangling block (\Cref{fig:tick_walkthrough}, Ticks~3,5). So the global pulses accumulate the $X$-error information onto every $A_1$. The $Z$-sub-cycle, which detects $Z$-errors via the $X$-stabilizer, follows a different pattern: Hadamards are applied to the A$_2$ ancillas only, so each A$_2$ acquires the $X$-eigenvalue of the surrounding data atoms. The opening and closing Hadamards are mutual inverses, and the OR-Toffoli$+$CCX block they sandwich is Clifford and commutes with every stabilizer. In the absence of errors, the whole syndrome-extraction-and-correction block therefore acts as the identity on $\bar X_j, \bar Z_j$.

\subsubsection{Error correction and ancilla reset}
\label{sec:correction_reset}
Once the syndrome is encoded, the correction step follows [\Cref{fig:model}(b)]: a second invocation of the same \textsc{ccx} gate on the same triple, now acting on syndrome-bearing ancillas (\Cref{fig:tick_walkthrough}, Tick~6) rather than raw ones, applies the Pauli correction, flipping the affected data qubit when both adjacent ancillas are in $\ket{1}$. Weight-one errors are healed in place; weight-$\geq 2$ events leave residual syndrome and rely on the code distance $d_{\rm code}=L$ for suppression.

Between consecutive correction sub-cycles, the addressed ancilla species is re-prepared in $\ket{0}$ by a species-selective optical-pumping pulse (shown in \Cref{fig:tick_walkthrough}, Tick~7), with duration set by the hyperfine-pumping cycle time of the ancilla species ($\sim\!1$--$5~\mu$s in current tweezer-array experiments~\cite{bluvstein2022quantum,singh2022dual,lis2023midcircuitoperationsusingomgarchitecture}). The coherent syndrome-extraction-and-correction block lasts only hundreds of nanoseconds, so this reset dominates the total per-sub-cycle wall-clock time.

\paragraph*{Worked example: a single $X$-error.}
Suppose an $X$-error fires on data qubit $D^\star$ in the toric-code ground state $\ket{\psi_0}$ (\Cref{fig:tick_walkthrough}, Ticks~1--2). Since $X$ anticommutes with $Z$, the two $B_p$ stabilizers containing $D^\star$ flip to $-1$ (pink), while their central ancilla qubits (A$_1$) are still in $\ket{0}$. 
The Hadamards of Tick~3 recast each (D, A$_1$, A$_1^\prime$) triple, where A$_1^\prime$ denotes the other A$_1$ ancilla bordering the same plaquette, as ancilla-target, so the \textsc{or-toffoli}\,+\,\textsc{ccx} sweeps (Tick~4) write the $Z$-parity of each $B_p$ onto its central A$_1$ via \Cref{eq:xor_factorization}. After the closing Hadamards (Tick~5), only the two A$_1$'s flanking $D^\star$ carry $\ket{1}$ (gold); all other A$_1$'s remain in $\ket{0}$. The correction \textsc{ccx} (Tick~6) then fires globally but, by its two-control logic, flips a data qubit only when \emph{both} adjacent A$_1$'s read $\ket{1}$---a condition uniquely met by the triple on $D^\star$, which is therefore flipped back while every other data qubit is untouched. A species-selective optical-pumping pulse resets the two excited A$_1$'s to $\ket{0}$ (Tick~7), restoring $\ket{\psi_0}$. The $Z$-sub-cycle is identical, with A$_2$ replacing A$_1$.

\subsection{Numerical simulation}
\label{sec:results}

Having established the gate-level building blocks, we proceed in three scales. First, we benchmark the per-triple unitaries under the full Lindblad model (\Cref{sec:gate_design}). Second, we verify the full pulse-level correction cycle on a 9-atom multi-triple patch using the time-dependent pulse Hamiltonian with details in Methods~\ref{app:pulse_cycle}. Third, we assess how the protocol performs on the full $4\!\times\!4$ rotated toric code with $n_d=16$ data qubits and $n_a=16$ ancillas (8 of each type), 32 sites in total.

\subsubsection{Per-triple gate fidelity}

\label{sec:gate_design}
\begin{figure}[!htbp]
    \centering
    \includegraphics[width=0.75\linewidth]{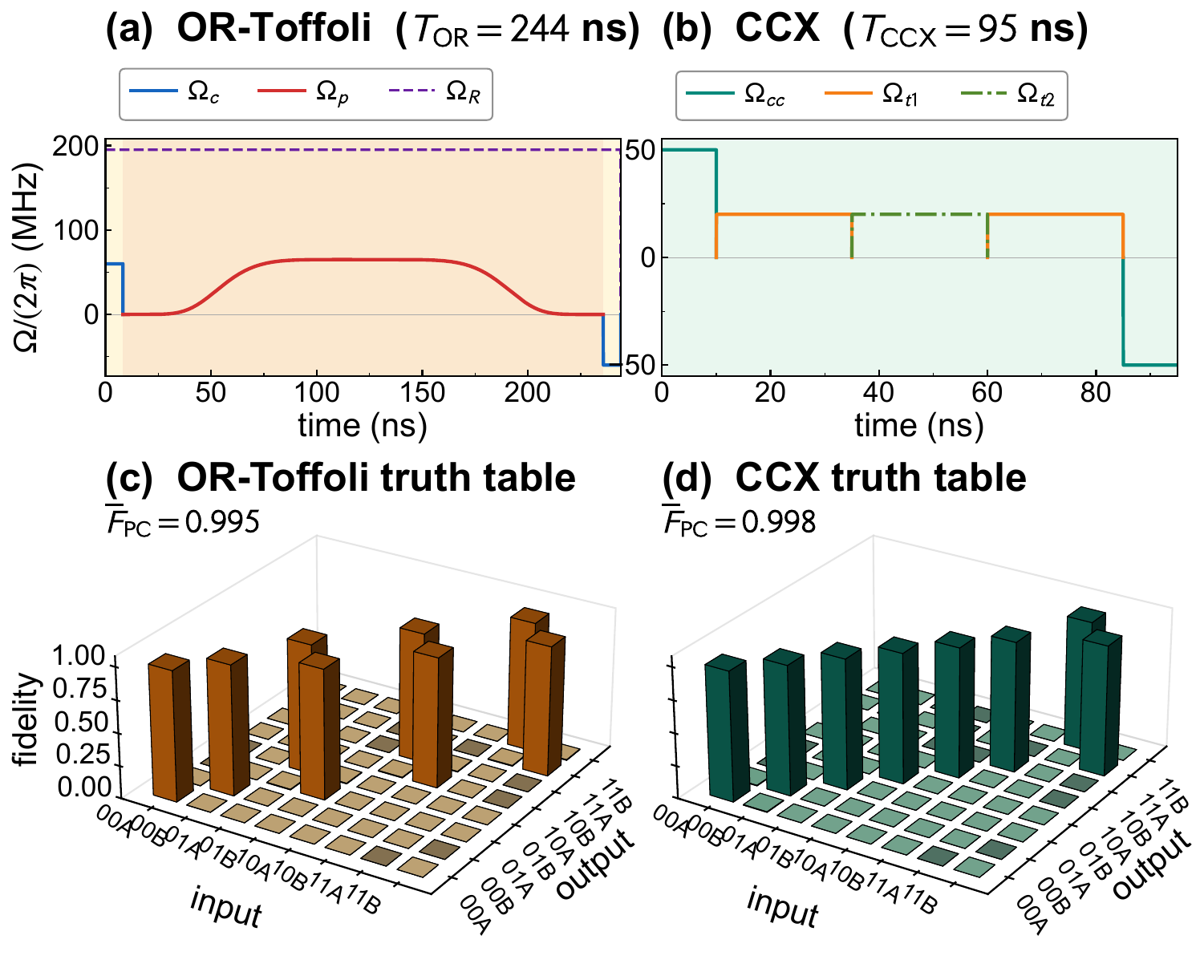}
    \caption{Pulse-level model and phase-corrected fidelity of the per-triple unitaries on a single (D, A$_1$, A$_1^\prime$) cell. State labels (subscript $a$ for ancilla, $d$ for data): ancilla ground states $\ket{0}_a,\ket{1}_a$ and Rydberg $\ket{r}_a$; data ground states $\ket{A}\!\equiv\!\ket{0}_d, \ket{B}\!\equiv\!\ket{1}_d$, intermediate $\ket{P}_d$, and Rydberg $\ket{R}_d$ (full level assignments in Methods, \Cref{app:pulse_details}). \textbf{(a)}~\textsc{or-toffoli} pulse sequence: two ancilla square $\pi$-pulses on $\ket{1}\!\leftrightarrow\!\ket{r}$ at $\Omega_c$ (blue) bracket a sixth-order super-Gaussian Raman drive $\Omega_p$ on the data $\ket{A,B}\!\leftrightarrow\!\ket{P}$ transition (red); the data $\ket{P}\!\leftrightarrow\!\ket{R}$ drive $\Omega_R$ (purple dashed) is held \emph{constant} throughout. \textbf{(b)}~\textsc{ccx} pulse sequence: two ancilla $\pi$-pulses on $\ket{0}\!\leftrightarrow\!\ket{r}$ at $\Omega_{cc}$ (teal) bracket three target sub-pulses on the data--Rydberg manifold:
    $\Omega_{t1}$ on $\ket{B}\!\leftrightarrow\!\ket{R}$ (orange solid) and $\Omega_{t2}$ on $\ket{A}\!\leftrightarrow\!\ket{R}$ (green dash-dot); $\ket{P}$ is never populated. \textbf{(c, d)}~Truth-table response of the per-triple unitary under the full Lindblad model of \Cref{app:pulse_verify}, for each of the eight computational-basis inputs $\ket{a_1 a_2 s}$ with ancilla bits $a_i\!\in\!\{0,1\}$ and data-qubit basis $s\!\in\!\{A,B\}$. Bar heights are output populations on the same 8-state basis. Pulse parameters and decoherence lifetimes are listed in \Cref{app:pulse_details,app:pulse_single}.}
    \label{fig:gate_fid_bar}
\end{figure}

Because local addressing is unavailable, the four-CNOT syndrome circuit is recompiled into global, species-selective pulses acting on the forced (D, A$_1$, A$_1^\prime$) triple, on which its XOR parity factorizes into exactly two sweeps: an \textsc{or-toffoli} and a \textsc{ccx} (\Cref{sec:syndrome_design}). These two global sweeps are the elementary operations of the whole protocol; we construct each at the pulse level on a single triple and benchmark it under a full open-system (Lindblad) model.

Both sweeps act on the three-atom level structure of \Cref{fig:model}(c): each ancilla (control) has two ground states $\ket{0}_a,\ket{1}_a$ and a Rydberg state $\ket{r}_a$, while the data atom (target) has two ground states $\ket{A}_d\!\equiv\!\ket{0}_d$ and $\ket{B}_d\!\equiv\!\ket{1}_d$, an intermediate state $\ket{P}_d$, and a Rydberg state $\ket{R}_d$ (specific atomic assignments in \Cref{app:pulse_details}). The two constructions use the Rydberg blockade in complementary ways. We realize the \textsc{or-toffoli} by electromagnetically induced transparency (EIT)~(\Cref{fig:gate_fid_bar}(a)). When both ancillas remain in $\ket{0}_a$, a sixth-order super-Gaussian Raman drive~(\Cref{eq:super_gaussian}) dresses the data atom through the intermediate state $\ket{P}_d$, which holds the data qubit in an EIT dark state that suppresses any flip. A single Rydberg-excited ancilla ($\ket{r}_a$) instead blockades $\ket{R}_d$ out of resonance, collapses the dark state, and turns the drive into a $\ket{A}_d\!\leftrightarrow\!\ket{B}_d$ rotation, so the data atom flips when at least one ancilla is excited.
 We realize the \textsc{ccx} directly from the blockade~(\Cref{fig:gate_fid_bar}(b)): three resonant sub-pulses on the $\ket{A,B}_d\!\leftrightarrow\!\ket{R}_d$ manifold compose to a full $\ket{A}_d\!\leftrightarrow\!\ket{B}_d$ inversion only when both ancillas are in $\ket{1}_a$ and the blockade is absent, while any excited ancilla detunes the sub-pulses and preserves the target. Skipping the slow two-photon Raman path makes the \textsc{ccx} sweep faster. Both are single global species-selective drives that act identically on every triple; pulse parameters and the rotating-frame Hamiltonians $H_{\rm OR/CCX}(t)$ are given in \Cref{app:pulse_details}.

We benchmark both sweeps by integrating the Lindblad master equation~\eqref{eq:lindblad} on the triple units. 
The model carries two physical error inputs: \textit{(i) spontaneous decay} of the three excited states, with lifetimes $\tau_r\!=\!138.9~\mu$s, $\tau_R\!=\!164.6~\mu$s, and $\tau_P\!=\!0.270~\mu$s drawn from the Alkali Rydberg Calculator (ARC) library~\cite{sibalic2017arc} (the intermediate $\ket{P}_d$ is three orders of magnitude shorter-lived than either Rydberg state); and \textit{(ii) residual same-species van der Waals coupling} between the two ancillas of the triple, captured by the ancilla--ancilla shift $V_{\rm cc}\,\ket{r}\!\bra{r}_{a_1}\!\otimes\!\ket{r}\!\bra{r}_{a_2}$ in $H_{\rm OR/CCX}$ and suppressed to $|V_{\rm cc}/V_{\rm ct}|\!\approx\!0.0063$ relative to the data--ancilla blockade $V_{\rm ct}$ (\Cref{app:interactions}). The post-calibration phase-corrected average gate fidelity~\cite{pedersen2007fidelity} [\Cref{eq:Fpc} of \Cref{app:pulse_single}], following the convention of Ref.~\cite{isenhower2010demonstration} for three-qubit Rydberg gates, is
\begin{equation}
    \bar F_{\rm PC}^{\,\textsc{ccx}} = 0.998, \qquad \bar F_{\rm PC}^{\,\textsc{or-toffoli}} = 0.995,
    \label{eq:Fpc_numerical}
\end{equation}
The hierarchy reflects two different bottlenecks. The \textsc{or-toffoli} ($T_{\rm OR}\!\approx\!244$~ns) is Rydberg-decay limited: the data and ancilla Rydberg states are populated for almost the entire two-photon Raman window, contributing ${\sim}60\%$ of the removable infidelity, with sub-leading $\ket{P}_d$-state decay and same-species interaction $V_{\rm cc}$. The \textsc{ccx} ($T_{\rm CCX}\!\approx\!95$~ns), with its shorter gate window, is limited mainly by the same-species $\ket{rr}_a$ van der Waals shift, with Rydberg decay sub-leading.
The full Lindblad operators and remaining simulation details are collected in \Cref{app:pulse_verify}.

\subsubsection{Full lattice verification}
\label{sec:multiround}
\label{sec:threshold}

For the third (code-level) scale, we simulate the full $4\!\times\!4$ correction cycle by representing the 32-qubit state as a matrix product state (MPS)~\cite{schollwock2011density,orus2014practical,vidal2003efficient,vidal2004efficient} and applying gates via tensor contraction with singular-value-decomposition (SVD) truncation.
Each run is initialized in the logical-zero state $\ket{00_L}$, the simultaneous $+1$ eigenstate of all stabilizers and of the two logical operators $\bar Z_1, \bar Z_2$, which we prepare deterministically.
The species-selective optical-pumping ancilla reset, the only non-unitary step in the correction round, is realized on the pure MPS by quantum-jump unraveling of $\gamma=1$ amplitude damping. The MPS ordering, truncation cutoff, ground-state preparation, and unraveling scheme are detailed in \Cref{app:mps,app:noise_full}.
\begin{figure}[!htbp]
    \centering
    \includegraphics[width=0.6\linewidth]{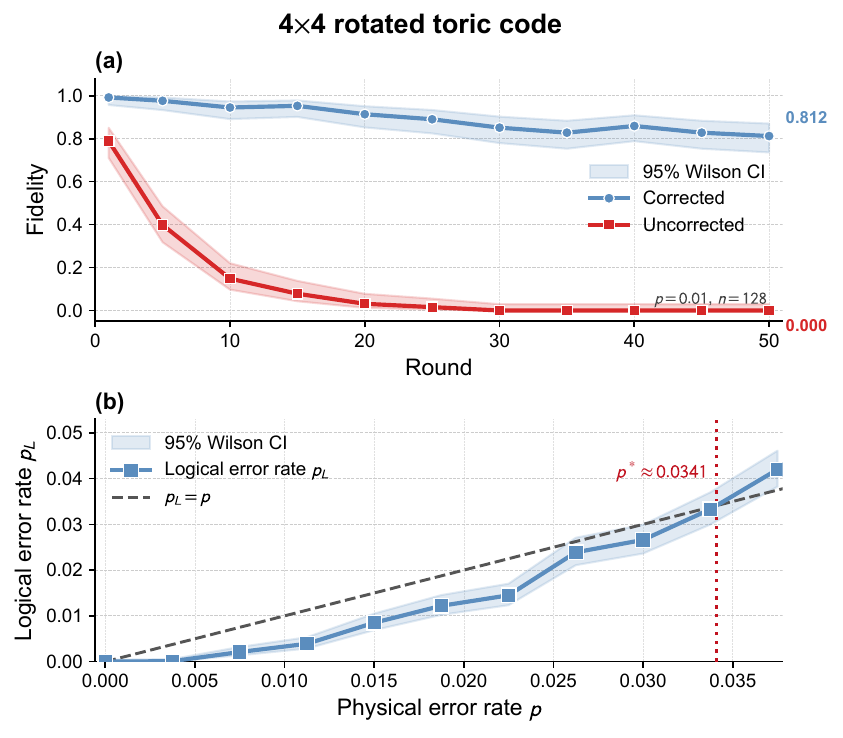}%
    \caption{Logical-memory benchmarks for the $4\!\times\!4$ rotated toric code.
    \textbf{(a)}~Multi-round fidelity evolution under i.i.d.\ depolarizing noise at physical error rate $p=0.01$ over $n=128$ independent Monte-Carlo trials. \emph{Blue circles:} corrected run --- every round applies the full measurement-free \textsc{or-toffoli}\,+\,\textsc{ccx} syndrome-extraction, correction, and projective ancilla-reset cycle, with the depolarizing channel of \Cref{eq:depolarizing} fired once at the start of each round; the mean fidelity decays gently from $\bar F\!\approx\!0.99$ at round $1$ to $\bar F\!\approx\!0.81$ at round $50$. \emph{Red squares:} uncorrected reference (no correction gates fire); the mean fidelity falls rapidly from $\bar F\!\approx\!0.79$ at round $1$ to statistical zero ($\bar F\!=\!0.000$) by round $\sim\!30$.
    \textbf{(b)}~Pseudo-threshold estimation. Logical error rate $p_L$ versus the physical depolarizing rate $p$; each data point averages $N_{\rm MC}=10000$ independent Monte-Carlo trials at $\chi_{\max}=256$. The black dashed line is the reference $p_L=p$; the red dotted vertical line marks the break-even crossing at $p^\star\!\approx\!0.0341$. The $p=0$ marker reports the $95\%$-confidence upper bound $p_L<3/N_{\rm MC}$ from the rule of three~\cite{eypasch1995probability}.
    Shaded bands in both panels are $95\%$ Wilson-score confidence intervals~\cite{wilson1927probable}. MPS parameters: maximum bond dimension $\chi_{\max}=96$ in~(a) and $\chi_{\max}=256$ in~(b), with truncation cutoff $\epsilon=10^{-14}$.}%
    \label{fig:ec_results}%
\end{figure}

We model the dominant code-level noise as an independent and identically distributed (i.i.d.)\ depolarizing channel applied once per correction round to each data qubit, before the syndrome-extraction sub-cycles fire (\Cref{sec:syndrome_design}; \Cref{fig:model}(b)):
\begin{equation}
    \mathcal{E}(\rho)=(1-p)\rho+\tfrac{p}{3}\bigl(X\rho X+Y\rho Y+Z\rho Z\bigr),
    \label{eq:depolarizing}
\end{equation}
where $p$ is a per-round error budget that lumps the calibrated gate-level channels of \Cref{sec:gate_design} into one phenomenological rate~\cite{dennis2002topological,fowler2012surface}; the scope of this surrogate is discussed in \Cref{sec:discussion}. One Monte-Carlo trial propagates $\ket{00_L}$ through $n_{\rm rounds}$ noisy correction rounds to a final pure state $\ket{\psi_{\rm corr}}$.

A working quantum memory must preserve logical information across many correction cycles, with a per-round logical error rate well below the unprotected baseline. We first probe this regime at a fixed physical error rate $p=0.01$, running $50$ correction cycles averaged over $n=128$ independent Monte-Carlo trials. Each trial tracks the state fidelity $F=|\braket{\psi_{\rm ref}|\psi_{\rm corr}}|^2$ against the noise-free reference $\ket{\psi_{\rm ref}}$ produced by the same prepared ground state. We write $\bar F$ for the Monte-Carlo mean of $F$ across independent trials; this is a \emph{state} fidelity, distinct from the Haar-averaged \emph{gate} fidelity $\bar F$ of \Cref{eq:pedersen_fidelity} used in Methods. \Cref{fig:ec_results} shows the result. With correction the fidelity drops gently and monotonically from $\bar F\!\approx\!0.99$ after the first round to $\bar F\!\approx\!0.81$ at round $50$. The uncorrected reference, by contrast, falls rapidly from $\bar F\!\approx\!0.79$ after one round and saturates at $\bar F=0.000$ (statistical zero) by round $\sim\!30$. Fitting the corrected curve to an exponential $\bar F(r)=(1-p_L)^r$ gives an effective per-round logical error rate $p_L\!\approx\!4.2\!\times\!10^{-3}$, versus $1-(1-p)^{n_d}\!\approx\!0.15$ per round for the uncorrected $n_d$-qubit memory ($n_d=16$).

We next sweep the physical error rate $p$ to locate the break-even point. At each $p$ we run $N_{\rm MC}=10000$ Monte-Carlo trials and record the empirical logical error rate $p_L$, defined as the fraction of trials whose recovered state fails to preserve the initial logical state~\cite{heussen2024measurementfree,Perlin_2023,acharya2024quantum}. \Cref{fig:ec_results}(b) shows the result. Below the break-even point the logical error rate sits well beneath the $p_L\!=\!p$ reference line; it crosses that line near $p\!\approx\!0.0341$. The crossing with the $p_L\!=\!p$ reference line yields a pseudo-threshold
\begin{equation}
    p^\star \approx 0.0341,
    \label{eq:threshold}
\end{equation}
below which the protocol actively suppresses errors and above which it introduces more errors than it removes. Because we have run only a single code size $L=4$, this is a pseudo-threshold in the proper sense: it marks the break-even point for that specific lattice, not a true scaling threshold extracted from finite-size scaling.

%%%%%%%%%%%%%%%%%%%%%%%%%%%%%%%%%%%%%%%%%%%%%%%%%%%%%%%%%%%%%%%%%%%%%%%%%%%%%%%
\section{Discussion}
\label{sec:discussion}
%%%%%%%%%%%%%%%%%%%%%%%%%%%%%%%%%%%%%%%%%%%%%%%%%%%%%%%%%%%%%%%%%%%%%%%%%%%%%%%
% npj QI: Discussion permits no subheadings, no "Limitations" or
% "Conclusions" subsections. Arc: (1) summary of the work, (2) advantages,
% (3) outlook framed as a roadmap. The italic lead-ins in the outlook
% paragraph are inline emphasis, not subsection headings.

We have proposed a toric-code quantum memory stabilized entirely by global, species-selective laser pulses, with no mid-circuit measurement, no atom transport, and no local addressing. A three-species Rydberg array carries the complete stabilizer cycle: syndrome extraction, coherent correction, and ancilla reset are all realized as global operations on (data, ancilla, ancilla) triples built from just two primitives, the \textsc{or-toffoli} and \textsc{ccx} sweeps. We tested the protocol at three scales: per-triple gate fidelity under a full Lindblad model, a pulse-level correction cycle on a nine-atom patch, and a code-level simulation of the $4\!\times\!4$ rotated code. The protocol extends the logical lifetime whenever the physical error rate stays below a pseudo-threshold $p^\star\approx 0.034$.

\begin{table*}[!htbp]
    \centering
    \caption{Comparison of measurement-based and measurement-free quantum error correction (QEC) rounds in neutral-atom arrays. Numbers are representative operating points from the cited literature; the key structural difference is that the measurement-free round avoids every operation with $\mu$s-scale overhead.}
    \label{tab:comparison}
    \renewcommand{\arraystretch}{1.4}
    \begin{tabular}{|l|c|c|}
        \hline
        \textbf{Feature} & \textbf{Measurement-based} & \textbf{This work} \\
        \hline\hline
        Syndrome readout & $\sim\!1$--$10$~ms & Coherent, $<\!1~\mu$s \\
        \hline
        Ancilla reset & re-preparation & Optical pumping, $\sim\!1$--$5~\mu$s \\
        \hline
        Atom shuttling & Required & Not required \\
        \hline
        Local addressing & Required & Not required \\
        \hline
        Classical feedback & Required & Not required \\
        \hline
        Atom species & 1 species & 3 species \\
        \hline
        Per-round atom loss & $\sim\!0.1\%$/move & Decay-limited \\
        \hline
        Cycle time & $\sim\!0.1$--$1$~ms & $\sim\!10~\mu$s \\
        \hline
    \end{tabular}
\end{table*}

The protocol eliminates every operation that incurs $\mu$s- to ms-scale overhead in a shuttle--measure--feedback architecture~\cite{bluvstein2022quantum,bluvstein2024logical}: mid-circuit readout, atom transport, and classical feedback (\Cref{tab:comparison}). The cycle time is therefore set by ancilla reset, which is one to two orders of magnitude shorter than a measurement-based round (\Cref{tab:comparison}). The cost is two extra ancilla species: a CSS code has two stabilizer types, and global species-selective control addresses each type with a separate species, so one data species and two ancilla species is the minimum. 

Three directions will sharpen and extend this result. \emph{Scale:} all numerics are at $L=4$, so $p^\star\approx 0.034$ is a break-even point rather than an asymptotic threshold; running the full correction circuit at $L=6,8$ is the immediate next step. \emph{Noise realism:} the code-level simulation applies an i.i.d.\ depolarizing channel to the data qubits only, so a circuit-level treatment that adds gate errors, ancilla-to-data error propagation, and erasure conversion of Rydberg decay~\cite{wu2022erasure,scholl2023erasure,ma2023high,cong2022hardware} is the natural refinement. \emph{Geometry:} the torus is not yet experimentally available, so planar variants with boundary ancillas or long-range-bonded surface codes~\cite{cong2022hardware} are the natural extension to a bounded geometry.

%%%%%%%%%%%%%%%%%%%%%%%%%%%%%%%%%%%%%%%%%%%%%%%%%%%%%%%%%%%%%%%%%%%%%%%%%%%%%%%
\section{Methods}
\label{sec:methods}
%%%%%%%%%%%%%%%%%%%%%%%%%%%%%%%%%%%%%%%%%%%%%%%%%%%%%%%%%%%%%%%%%%%%%%%%%%%%%%%
% npj QI: All Methods must be in the main file (not Supplementary).
% Former appendices A, B, C are demoted to subsections here.

\subsection{Rydberg interaction details}
\label{app:interactions}

This section collects the Rydberg interaction parameters used in the pulse design of \Cref{sec:gate_design}. The species assignments ($^{87}$Rb, $^{133}$Cs, $^{39}$K) and the rationale for using three species are given in \Cref{sec:scheme-overview,sec:syndrome_design}; here we specify the F\"orster pair channels and quantify the selectivity hierarchy.

The key design requirement is that the interspecies data--ancilla coupling $V_{\rm dd}$ must be strong enough to produce a clean Rydberg blockade, while the same-species residual couplings $V_{\rm vdW}^{(\rm AA)}$ \emph{and} $V_{\rm vdW}^{(\rm DD)}$ must both remain weak enough not to spoil the gate logic when neighbouring atoms of one species are excited simultaneously by a global pulse. This is achieved by choosing $d$-state F\"orster resonances, which provide strong resonant dipole--dipole interspecies coupling ($\propto 1/r^3$) while keeping the intraspecies van der Waals interaction ($\propto 1/r^6$) suppressed on both same-species channels simultaneously.

In the rotated toric-code geometry of \Cref{fig:model}(a), data qubits sit on lattice vertices and the two ancilla species occupy alternating plaquette centers in a checkerboard pattern. The nearest data--ancilla pair sits at distance $r_{\rm DA}=a/\sqrt{2}$ (diagonal across half a plaquette), the nearest same-species ancilla pair at $r_{\rm AA}=a\sqrt{2}$ (diagonal across one checkerboard step), and the nearest data--data pair at $r_{\rm DD}=a$ (one lattice step along a Cartesian axis). With this geometry, the two same-species selectivity ratios read
\begin{align}
    \frac{V_{\rm dd}}{V_{\rm vdW}^{(\rm AA)}}
    &= \frac{\tilde C_3/r_{\rm DA}^3}{|C_6^{(\rm AA)}|/r_{\rm AA}^6}
    = \frac{16\sqrt{2}\,\tilde C_3}{|C_6^{(\rm AA)}|}\,a^3, \notag\\
    \frac{V_{\rm dd}}{V_{\rm vdW}^{(\rm DD)}}
    &= \frac{\tilde C_3/r_{\rm DA}^3}{|C_6^{(\rm DD)}|/r_{\rm DD}^6}
    = \frac{2\sqrt{2}\,\tilde C_3}{|C_6^{(\rm DD)}|}\,a^3,
    \label{eq:interaction_ratio}
\end{align}
where $\tilde C_3$ denotes the effective resonant dipole--dipole coefficient for the chosen F\"orster pair, evaluated for pair vectors in the lattice plane with the quantization axis perpendicular to it. 
Each ratio in \Cref{eq:interaction_ratio} is evaluated with the coefficients of the active sub-cycle: $\tilde C_3$ and $C_6^{(\rm AA)}$ are coefficients of Rb--Cs and Cs--Cs interaction strengths when A$_1$ is driven, and are Rb--K and K--K interaction strengths when A$_2$ is driven, while $C_6^{(\rm DD)}$ is the Rb--Rb interaction strength coefficient throughout.
Both ratios scale as $a^3$, so selectivity improves with larger lattice spacing on both channels under certain coefficients. 

For the D--A$_1$ (Rb--Cs) channel, we use the near-resonant F\"orster pair $\ket{54D_{3/2};62D_{5/2}}$~\cite{ireland2024interspecies}. For the D--A$_2$ (Rb--K) channel, we use the resonant F\"orster pair $\ket{54D_{3/2};51D_{3/2}}\to\ket{55P_{3/2};51F_{5/2}}$, sharing the channel structure analyzed in Ref.~\cite{otto2020interspecies}. The resulting interaction coefficients and blockade strengths at lattice spacing $a=4~\mu$m are collected in \Cref{tab:interactions_X,tab:interactions_Z}; same-species van der Waals coefficients are computed with ARC~\cite{sibalic2017arc}. 
The same-species selectivity ratios listed in \Cref{tab:interactions_X,tab:interactions_Z} all sit comfortably above the strong-blockade threshold $\gtrsim 10^2$, so simultaneously Rydberg-excited same-species atoms remain effectively non-interacting on the gate time-scale.

\begin{table*}[!htbp]
\caption{$X$-round Rydberg interaction channels at lattice spacing $a=4~\mu$m: data D~=~Rb with $X$-round ancilla A$_1$~=~Cs. The interspecies D--A$_1$ pair interacts via resonant dipole--dipole coupling ($\tilde C_3/r^3$); same-species pairs (A$_1$--A$_1$ and D--D) via van der Waals (vdW) ($C_6/r^6$). The last column reports the strong-blockade selectivity ratio $V_{\rm dd}/V_{\rm vdW}$ from \Cref{eq:interaction_ratio}.}
\label{tab:interactions_X}
\centering
\renewcommand{\arraystretch}{1.4}
\begin{tabular}{|l|l|l|c|c|}
\hline
\textbf{Channel} & \textbf{Role} & \textbf{Coefficient} & \textbf{Separation} & \textbf{$V_{\rm dd}/V_{\rm vdW}$} \\
\hline\hline
 Rb--Cs & D--A$_1$ (inter.)   & $\tilde C_3=10.0$~GHz$\cdot\mu$m$^3$ & $a/\sqrt{2}$ & ---   \\
\hline
 Cs--Cs & A$_1$--A$_1$ (same) & $C_6=-91.0$~GHz$\cdot\mu$m$^6$       & $a\sqrt{2}$  & $159$ \\
\hline
 Rb--Rb & D--D (same)         & $C_6=+9.07$~GHz$\cdot\mu$m$^6$       & $a$          & $200$ \\
\hline
\end{tabular}
\end{table*}

\begin{table*}[!htbp]
\caption{$Z$-round Rydberg interaction channels at $a=4~\mu$m: data D~=~Rb with $Z$-round ancilla A$_2$~=~K. Channel types, the selectivity ratio $V_{\rm dd}/V_{\rm vdW}$, and the numeric separations follow \Cref{tab:interactions_X}.}
\label{tab:interactions_Z}
\centering
\renewcommand{\arraystretch}{1.4}
\begin{tabular}{|l|l|l|c|c|}
\hline
\textbf{Channel} & \textbf{Role} & \textbf{Coefficient} & \textbf{Separation} & \textbf{$V_{\rm dd}/V_{\rm vdW}$} \\
\hline\hline
 Rb--K  & D--A$_2$ (inter.)   & $\tilde C_3\approx 8.3$~GHz$\cdot\mu$m$^3$ & $a/\sqrt{2}$ & ---    \\
\hline
 K--K   & A$_2$--A$_2$ (same) & $C_6\approx-9.9$~GHz$\cdot\mu$m$^6$        & $a\sqrt{2}$  & $1200$ \\
\hline
 Rb--Rb & D--D (same)         & $C_6=+9.07$~GHz$\cdot\mu$m$^6$             & $a$          & $166$  \\
\hline
\end{tabular}
\end{table*}

\subsection{Pulse-level verification}
\label{app:pulse_verify}

This section validates the protocol at the pulse level in two complementary ways. Methods~\ref{app:pulse_single} solves the Lindblad master equation for the individual three-qubit units, identifying the dominant error channels. Methods~\ref{app:pulse_cycle} runs the X-error sub-cycle on a nine-atom unit cell with purely coherent evolution. On representative computational-basis inputs, the pulse sequence reproduces the gate-circuit logic of \Cref{sec:syndrome_design}, validating the pulse compilation.

\subsubsection{Pulse details for the three-qubit unitaries}
\label{app:pulse_details}

We list the atomic levels, pulse parameters, and Hamiltonians for the \textsc{or-toffoli} and \textsc{ccx} sweeps used in \Cref{sec:gate_design}, taking the D--A$_1$ pair as the worked example; the species and interaction parameters are listed in Methods~\ref{app:interactions}. The physical construction of the two sweeps (EIT dark state plus blockade for the \textsc{or-toffoli}, direct resonant blockade for the \textsc{ccx}) is given in \Cref{sec:gate_design}; the level structure is shown in \Cref{fig:model}(c) and the pulse sequences in \Cref{fig:gate_fid_bar}(a,~b). Specific atomic-state assignments and pulse parameters are given below.

The \textsc{or-toffoli} sweep (introduced in \Cref{sec:syndrome_design}) proceeds in three stages. First, a square $\pi$-pulse at Rabi frequency $\Omega_c$ excites the ancilla qubits from $\ket{1}_a$ to the Rydberg state $\ket{r}_a$. Next, a two-photon Raman pulse on data couples $\ket{A}_d/\ket{B}_d\leftrightarrow\ket{P}_d$ with super-Gaussian envelope $\Omega_p(t)$ and large detuning $\Delta$, while $\ket{P}_d\leftrightarrow\ket{R}_d$ is continuously driven by $\Omega_R$. When both ancilla qubits in the three-atom unit remain in $\ket{0}_a$, destructive interference between the two pathways forms the EIT dark state with no state flip on the data qubit; when at least one ancilla qubit is in $\ket{r}_a$, the blockade shifts $\ket{R}_d$ far out of resonance, collapses the dark state, and converts the pulse into an effective $\ket{A}_d\leftrightarrow\ket{B}_d$ Raman $\pi$-rotation. Finally, a second $\pi$-pulse on the ancilla qubits restores $\ket{r}_a\to\ket{1}_a$.

The \textsc{ccx} sweep is designed directly from the Rydberg blockade. First, a $\pi$-pulse drives $\ket{0}_a\to\ket{r}_a$ for the ancilla qubits, so that blockade is active whenever any ancilla qubit is in $\ket{0}_a$. The operation on the data qubit consists of three sequential sub-pulses: $\ket{B}_d\leftrightarrow\ket{R}_d$ (Rabi frequency $\Omega_{t1}$), $\ket{A}_d\leftrightarrow\ket{R}_d$ ($\Omega_{t2}$), and $\ket{B}_d\leftrightarrow\ket{R}_d$ again. When any control occupies $\ket{r}_a$, the blockade renders all three sub-pulses off-resonant and the target is preserved; when both controls remain in $\ket{1}_a$ (ground state, no blockade), the sub-pulses compose to a complete $\ket{A}_d\leftrightarrow\ket{B}_d$ inversion. A final $\pi$-pulse restores $\ket{r}_a\to\ket{0}_a$. The total gate time is $T_{\rm CCX}=2T_{cc}+3T_t\approx 95$~ns, roughly $2.6\times$ faster than the \textsc{or-toffoli} sweep because the \textsc{ccx} does not rely on a slow two-photon Raman process through an intermediate state.

\paragraph{Atomic states.} Throughout this section, subscript $a$ labels an ancilla register and subscript $d$ labels the data register. Each ancilla qubit (Cs) has two ground states $\ket{0}_a = \ket{6S_{1/2},F\!=\!3}$, $\ket{1}_a = \ket{6S_{1/2},F\!=\!4}$ and one Rydberg state $\ket{r}_a=\ket{62D_{5/2},m_j\!=\!+5/2}$. Each data qubit (Rb) has two ground states $\ket{A}_d\equiv\ket{0}_d=\ket{5S_{1/2},F\!=\!1}$, $\ket{B}_d\equiv\ket{1}_d=\ket{5S_{1/2},F\!=\!2}$, an intermediate state $\ket{P}_d=\ket{7P_{3/2}}$, and a Rydberg state $\ket{R}_d=\ket{54D_{3/2},m_j\!=\!+3/2}$.

\paragraph{Pulse parameters.} The two-photon target drive of the \textsc{or-toffoli} sweep uses a sixth-order super-Gaussian envelope,
\begin{equation}
    \Omega_p(t)=\frac{\Omega_p}{2}\exp\!\Bigl[-\Bigl(\frac{t-t_c}{\tau_w}\Bigr)^{6}\Bigr],\quad t\in[T_c,\,T_c+2T_f],
    \label{eq:super_gaussian}
\end{equation}
with window half-width $T_f$, $1/e$ half-width $\tau_w$, and centre $t_c=T_c+T_f$, chosen so that the two-photon pulse area $\int\Omega_p^2(t)/(2\Delta)\,dt=\pi$. The remaining drives ($\Omega_c$, $\Omega_{cc}$, $\Omega_{t1}$, $\Omega_{t2}$) are square pulses. For the \textsc{or-toffoli} sweep we use $\Omega_p=2\pi\!\times\!65$~MHz, $\Omega_R=3.0\,\Omega_p=2\pi\!\times\!195$~MHz, $\Omega_c=2\pi\!\times\!60$~MHz, $\Delta=2\pi\!\times\!500$~MHz, control $\pi$-pulse duration $T_c=\pi/\Omega_c\approx 8.33$~ns, super-Gaussian target half-window $T_f\approx 113.7$~ns (order~6, canonical $\pi/4$ two-photon area), giving total gate time $T_{\rm OR}=2T_c+2T_f\approx 244$~ns. For the \textsc{ccx} sweep: $\Omega_{cc}=2\pi\!\times\!50$~MHz, $\Omega_t=2\pi\!\times\!20$~MHz, $T_{cc}=\pi/\Omega_{cc}=10$~ns, $T_t=\pi/\Omega_t=25$~ns, total gate time $T_{\rm CCX}=2T_{cc}+3T_t=95$~ns. Both pulse sequences are plotted in \Cref{fig:gate_fid_bar}(a,~b), and the resulting phase-corrected fidelities $\bar F_{\rm PC}$ are reported in \Cref{sec:gate_design}.

\subsubsection{Single-gate simulation}
\label{app:pulse_single}

We validate the fidelity of different three-qubit units by solving the Lindblad master equation for the full three-atom open system (two ancilla qubits (Cs) as controls and one data qubit (Rb) as the target). The density matrix $\rho$ of the three-atom system evolves under
\begin{equation}
    \dot\rho = -i\bigl[H(t),\,\rho\bigr] + \sum_k \gamma_k \Bigl(L_k\,\rho\,L_k^\dagger - \tfrac{1}{2}\bigl\{L_k^\dagger L_k,\,\rho\bigr\}\Bigr),
    \label{eq:lindblad}
\end{equation}
where $H(t)$ is the time-dependent Hamiltonian specified below, $L_k=\ket{g_k}\!\bra{e_k}$ are collapse operators describing spontaneous decay from excited state $\ket{e_k}$($\ket{P}$, $\ket{r}$, $\ket{R}$) to ground state $\ket{g_k}$, and $\gamma_k=1/\tau_k$ are the corresponding decay rates. The equation is integrated numerically using the \texttt{mesolve} solver in QuTiP~\cite{johansson2012qutip}. The average gate fidelities reported in \Cref{sec:gate_design} are obtained from these simulations.

\paragraph{Hamiltonian.} The three-atom Hamiltonian is written in the rotating frame under the rotating-wave approximation. For the \textsc{or-toffoli} sweep,
\begin{align}
    H_{\rm OR}(t) &= \Delta\ket{P}\!\bra{P} \notag\\
    &+ V_{\rm ct}\bigl(\ket{r}\!\bra{r}_{a_1}\!+\!\ket{r}\!\bra{r}_{a_2}\bigr)\!\otimes\!\ket{R}\!\bra{R} \notag\\
    &+ V_{\rm cc}\,\ket{r}\!\bra{r}_{a_1}\!\otimes\!\ket{r}\!\bra{r}_{a_2} \notag\\
    &+ \tfrac{\Omega_c(t)}{2}\bigl(\ket{1}\!\bra{r}_{a_1}\!+\!\ket{1}\!\bra{r}_{a_2}\!+\!\mathrm{H.c.}\bigr) \notag\\
    &+ \tfrac{\Omega_p(t)}{2}\bigl(\ket{A}\!\bra{P}\!+\!\ket{B}\!\bra{P}\!+\!\mathrm{H.c.}\bigr) \notag\\
    &+ \tfrac{\Omega_R}{2}\bigl(\ket{P}\!\bra{R}\!+\!\mathrm{H.c.}\bigr),
    \label{eq:H_OR}
\end{align}
where the first line is the intermediate-state detuning, the second and third lines are the interspecies blockade and intraspecies van der Waals interactions, and the last three lines are the laser--atom couplings (control $\pi$-pulses, two-photon probe, and Rydberg coupling). For the \textsc{ccx} gate the Hamiltonian has the same interaction terms but different drive structure: the ancilla qubits are excited on the opposite branch $\ket{0}\!\leftrightarrow\!\ket{r}$ (so blockade is active when any ancilla is $\ket{0}$), there is no intermediate-state detuning, and the target is driven directly by sequential sub-pulses $\ket{B}\!\leftrightarrow\!\ket{R}$ and $\ket{A}\!\leftrightarrow\!\ket{R}$:
\begin{align}
    H_{\rm CCX}(t) &= V_{\rm ct}\bigl(\ket{r}\!\bra{r}_{a_1}\!+\!\ket{r}\!\bra{r}_{a_2}\bigr)\!\otimes\!\ket{R}\!\bra{R} \notag\\
    &+ V_{\rm cc}\,\ket{r}\!\bra{r}_{a_1}\!\otimes\!\ket{r}\!\bra{r}_{a_2} \notag\\
    &+ \tfrac{\Omega_{cc}(t)}{2}\bigl(\ket{0}\!\bra{r}_{a_1}\!+\!\ket{0}\!\bra{r}_{a_2}\!+\!\mathrm{H.c.}\bigr) \notag\\
    &+ \tfrac{\Omega_{t1}(t)}{2}\bigl(\ket{B}\!\bra{R}\!+\!\mathrm{H.c.}\bigr) \notag\\
    &+ \tfrac{\Omega_{t2}(t)}{2}\bigl(\ket{A}\!\bra{R}\!+\!\mathrm{H.c.}\bigr).
    \label{eq:H_CCX}
\end{align}
The interaction strengths $V_{\rm ct}/(2\pi)=+441.9$~MHz and $V_{\rm cc}/(2\pi)=-2.78$~MHz are derived from the $\tilde C_3$ and $C_6$ coefficients in Methods~\ref{app:interactions} at the lattice spacing $a=4~\mu$m. 

\paragraph{Collapse operators.} Three excited states undergo spontaneous decay. For each, we include two collapse operators $L_k=\sqrt{\gamma_k/2}\,\ket{g_k}\!\bra{e_k}$, one per ground state, with equal branching ratios:
\begin{itemize}
    \item Cs Rydberg: $\ket{r}_a\!=\!\ket{62D_{5/2}}\!\to\!\ket{0}_a,\,\ket{1}_a$ \;($\tau_r=138.9~\mu$s),
    \item Rb Rydberg: $\ket{R}_d\!=\!\ket{54D_{3/2}}\!\to\!\ket{A}_d,\,\ket{B}_d$ \;($\tau_R=164.6~\mu$s),
    \item Rb intermediate: $\ket{P}_d\!=\!\ket{7P_{3/2}}\!\to\!\ket{A}_d,\,\ket{B}_d$ \;($\tau_P=0.270~\mu$s).
\end{itemize}
The dominant error source for the \textsc{or-toffoli} sweep is Rydberg-state decay accumulated over the long gate window $T_{\rm OR}\!\approx\!244$~ns (the back-of-envelope $\gamma_R\,T_f\!\approx\!1.4\!\times\!10^{-3}$ matches the channel-resolved simulation contribution of $1.7\!\times\!10^{-3}$). The intermediate-state $\ket{P}_d$ and the same-species $V_{\rm cc}$ shift on the $\ket{rr}_a$ branch contributes to the left. 
By contrast, the \textsc{ccx} sweep is dominated by $V_{\rm cc}$ on the $3T_t\!=\!75$~ns $\ket{rr}_a$-parked window, with Rydberg decay sub-leading and $\ket{P}_d$ decay structurally absent.

\paragraph{Fidelity computation.} We benchmark each gate by the Haar-averaged average gate fidelity~\cite{pedersen2007fidelity},
\begin{equation}
    \bar F(\mathcal E,\,U_0) \;=\; \frac{\mathcal{D}\,F_{\rm pro}(\mathcal E,\,U_0)+1}{\mathcal{D}+1},
    \label{eq:pedersen_fidelity}
\end{equation}
where $\mathcal E$ is the simulated completely positive, trace-preserving (CPTP) gate, $U_0$ the ideal target unitary, and the process (entanglement) fidelity on the $\mathcal{D}=2^n$-dimensional computational subspace is
\begin{equation}
    F_{\rm pro}(\mathcal E,\,U_0) \;=\; \frac{1}{\mathcal{D}^{2}}\sum_{i,j=1}^{\mathcal{D}}\bra{\pi(i)}\,\mathcal E(\rho_{ij})\,\ket{\pi(j)},
    \label{eq:Fpro}
\end{equation}
with $\rho_{ij}=\ket{\psi_i}\!\bra{\psi_j}$ the computational-basis operators, $U_0\in\{U_\textsc{or-toffoli}, U_\textsc{ccx}\}$ the ideal unit-cell target [\Cref{eq:xor_factorization}], and $\ket{\pi(i)}=U_0\ket{\psi_i}$ the ideal outputs; for both three-qubit gates here $n=3$ and $\mathcal{D}=8$. $F_{\rm pro}$ is obtained from the rank-four Choi tensor of $\mathcal E$ restricted to the $8$-dimensional computational subspace, computed via $\mathcal{D}^{2}=64$ Lindblad runs of the $36$-dimensional three-atom simulator with non-Hermitian initial operators $\rho_{ij}$. Following the unit-cell decomposition of \Cref{sec:syndrome_design}, verified coherently on nine atoms in Methods~\ref{app:pulse_cycle}, this isolated-triple channel represents the per-cell channel of the full-lattice global pulse to within numerical accuracy at the design spacing $a=4~\mu$m; lattice-scale corrections are subsumed into the depolarizing channel of \Cref{eq:depolarizing}.

The bare fidelity~\eqref{eq:pedersen_fidelity} penalises every coherent diagonal phase the gate writes onto the computational basis. The single-qubit-$Z$ components are absorbed by per-species pulse-phase calibration at zero pulse cost; following the convention of Ref.~\cite{isenhower2010demonstration} for three-qubit Rydberg gates, we report the phase-corrected average gate fidelity
\begin{equation}
    \bar F_{\rm PC} \;=\; \max_{\substack{D\,\text{diagonal}\\ |D_{kk}|=1}}\; \bar F\bigl(\mathcal E,\,D\,U_0\bigr),
    \label{eq:Fpc}
\end{equation}
maximised over all $8\!\times\!8$ diagonal unitaries $D$ on the computational basis. The resulting numerical fidelities and truth-table matrices are reported in \Cref{sec:gate_design} and \Cref{fig:gate_fid_bar}.

\subsubsection{Unit-cell verification}
\label{app:pulse_cycle}

To verify that the composite pulse sequence implements the intended correction logic, we replace each gate in the circuit of \Cref{fig:tick_walkthrough} with its Rydberg pulse sequence from \Cref{sec:gate_design} and propagate the nine-atom state under the full time-dependent Hamiltonian (coherent Schr\"odinger evolution, without Lindblad decay). A full Lindblad treatment of the nine-atom Hilbert space at the pulse level is computationally prohibitive; the decoherence effects on the individual gates are instead captured in Methods~\ref{app:pulse_single}. The unit cell consists of four data qubits ($D_2,D_3,D_5,D_6$) and five A$_1$ ancillas ($a,b,c,d,e$) with periodic boundary conditions. We use a 5-index ket $\ket{D_2\,D_3\,D_5\,D_6\,a}$ to label the state of the four data qubits and the central ancilla $a$; the peripheral ancillas $b,c,d,e$ are tracked internally but not shown in the ket labels, since they all return to $\ket{0}$ at the end of each sub-cycle.

\begin{figure*}[!htbp]
    \centering
    \includegraphics[width=0.95\linewidth]{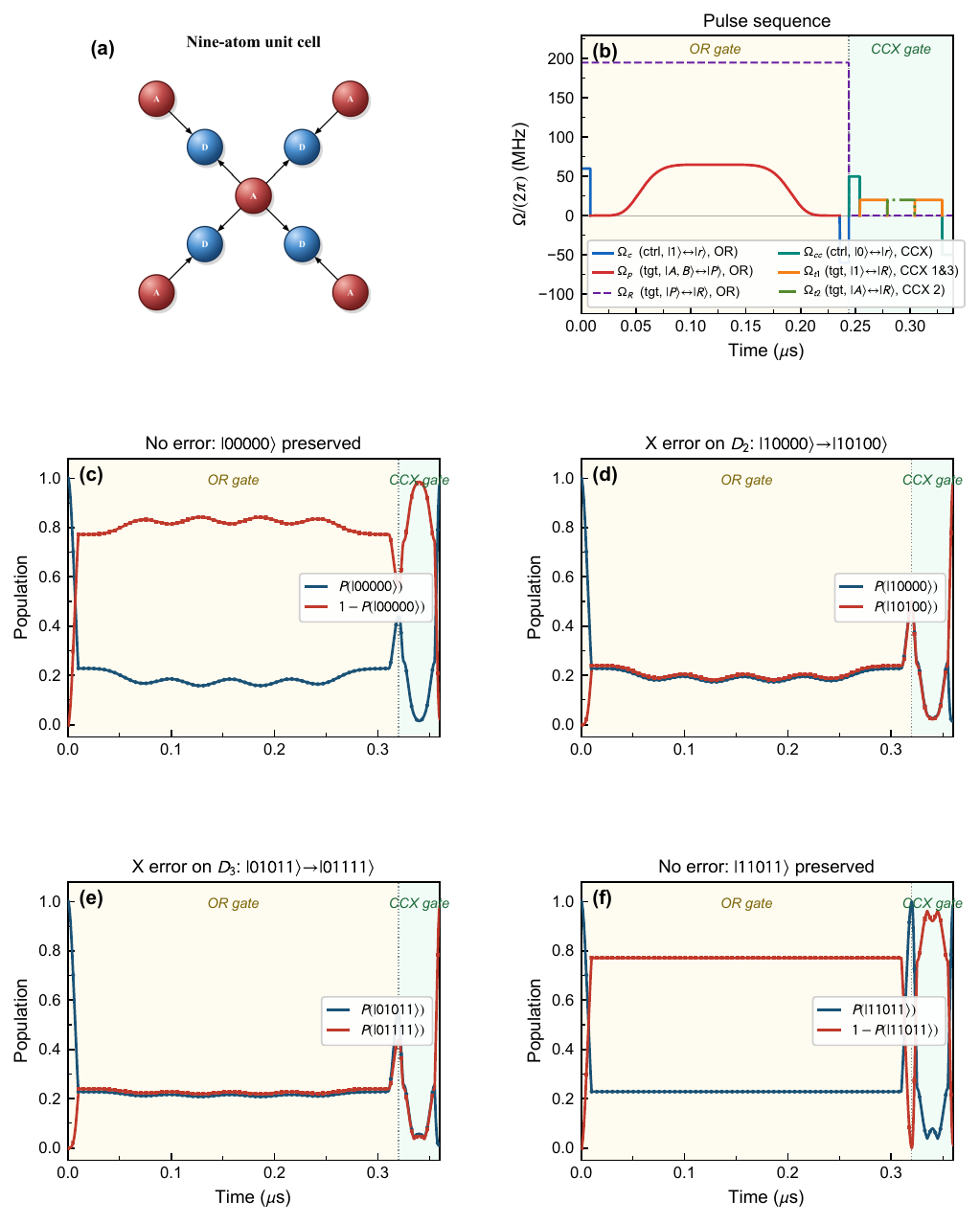}
    \caption{Pulse-level unitary simulation of the complete $X$-error sub-cycle on the nine-atom unit cell (coherent Hamiltonian evolution, no Lindblad decay). \textbf{(a)}~Nine-atom unit cell used in the verification: four data qubits (D) and five A$_1$ ancillas (A) with the lattice adjacency of the rotated toric code. \textbf{(b)}~Pulse envelopes for all driving fields during one complete \textsc{or-toffoli}+\textsc{ccx} cycle at the lattice-spacing $a=4~\mu$m operating point of \Cref{app:pulse_details} ($T_{\rm OR}\!\approx\!244$~ns, $T_{\rm CCX}\!=\!95$~ns); yellow shading marks the \textsc{or-toffoli} stage, green shading the \textsc{ccx} stage. \textbf{(c,f)}~No-error inputs: $\ket{00000}$ and $\ket{11011}$ are returned to their initial states after the full cycle, confirming that the protocol leaves error-free states undisturbed. \textbf{(d,e)}~Single-$X$-error inputs: an $X$-error on $D_2$ transfers $\ket{10000}\to\ket{10100}$, and an $X$-error on $D_3$ transfers $\ket{01011}\to\ket{01111}$, demonstrating correct syndrome extraction and coherent correction. In all population panels, blue curves show the initial-state population and red curves the target-state population. Panels (c)--(f) were simulated at the previous design point ($T_{\rm OR}\!=\!320$~ns, $T_{\rm CCX}\!=\!40$~ns), which is why their time axis extends slightly past panel~(b); the unit-cell logic-check conclusions are independent of these gate-time choices.}
    \label{fig:pulse_sim}
\end{figure*}

\Cref{fig:pulse_sim} shows the simulation results for four representative initial states. Panel~(a) illustrates the nine-atom unit cell and panel~(b) shows the pulse envelopes for all driving fields during one complete \textsc{or-toffoli}+\textsc{ccx} cycle, with yellow shading marking the \textsc{or-toffoli} stage and green shading the \textsc{ccx} stage.

\textit{Error-free inputs.} For the initial states $\ket{00000}$ [panel~(c)] and $\ket{11011}$ [panel~(f)], the population returns to the initial state after the complete cycle, confirming that the protocol correctly identifies the absence of errors and leaves the data qubits undisturbed.

\textit{Single $X$-error inputs.} Panels~(d) and~(e) demonstrate error detection and correction. An $X$-error on $D_2$ [panel~(d)] triggers the state transfer $\ket{10000}\to\ket{10100}$: the \textsc{or-toffoli} stage maps the error syndrome onto the adjacent A$_1$ ancillas, and the \textsc{ccx} stage applies the conditional correction, flipping the ancilla and restoring the data. Similarly, an $X$-error on $D_3$ [panel~(e)] produces $\ket{01011}\to\ket{01111}$. In both cases the population is cleanly transferred from the initial (blue) to the target (red) state by the end of the cycle.

The four computational-basis test cases confirm the protocol logic on product-state inputs: error-free states are preserved and single-qubit errors are corrected by the composite pulse sequence. Since this is a coherent Schr\"odinger simulation without Lindblad decay, the results validate the protocol logic independently of the decoherence effects quantified at the single-gate level in Methods~\ref{app:pulse_single}.

\subsection{Code-level simulation: MPS framework and noise model}
\label{app:simulation}

This section collects the technical details of the matrix-product-state (MPS) circuit simulator and the Monte-Carlo unraveling used to produce the multi-round and pseudo-threshold data of \Cref{sec:multiround,sec:threshold}.

\subsubsection{MPS framework}
\label{app:mps}

We represent the $32$-qubit state as a matrix product state~\cite{schollwock2011density,orus2014practical,vidal2003efficient,vidal2004efficient},
\begin{equation}
    \ket{\Psi} = \sum_{s_1,\ldots,s_N} M^{s_1}_1 M^{s_2}_2 \cdots M^{s_N}_N \ket{s_1,\ldots,s_N},
\end{equation}
where $N=32$ is the number of sites, $s_i\in\{0,1\}$ is the physical index of qubit $i$ and each $M^{s_i}_i$ is a $\chi_{i-1}\!\times\!\chi_i$ complex matrix, with open boundary conditions $\chi_0=\chi_N=1$ so that the matrix product evaluates to a scalar amplitude. Each bond dimension $\chi_i$ upper-bounds the Schmidt rank of $\ket{\Psi}$ across the bipartition between sites $\{1,\ldots,i\}$ and $\{i+1,\ldots,N\}$. The achieved maximum bond dimension $\chi\equiv\max_i\chi_i$ therefore sets the entanglement the ansatz can faithfully represent, and capping it at a per-run ceiling $\chi_{\max}$ is the single truncation knob exposed by the simulator.

Gates are applied via local SVD truncation, the standard MPS circuit-simulation primitive~\cite{vidal2003efficient,vidal2004efficient,schollwock2011density}. A local gate on consecutive sites $i,\ldots,i+w-1$ ($w=2,3$ in our protocol) is absorbed into the $w$-site tensor $\Theta^{s_i\cdots s_{i+w-1}}$ and split back into $w$ MPS tensors by a cascade of $w-1$ SVDs, with three-site contractions used for the \textsc{or-toffoli} and \textsc{ccx} sweeps. On each cut we retain the leading $\chi_{\rm cut}$ singular values $\sigma_k$, choosing the smallest $\chi_{\rm cut}$ whose discarded tail weight $\sum_{k>\chi_{\rm cut}}\sigma_k^2/\sum_k\sigma_k^2$ falls below $\epsilon=10^{-14}$, subject to the per-run cap $\chi_{\rm cut}\leq\chi_{\max}$ (\Cref{sec:multiround,sec:threshold}). The implementation uses the \texttt{quimb} \texttt{CircuitMPS} backend~\cite{gray2018quimb}.

We map the 2D lattice to the 1D MPS chain by interleaving columns of data qubits (even columns) with columns of ancilla qubits (odd columns), each column scanned from $y=0$ to $y=L-1$; this keeps data qubits and their nearest ancillas close in the 1D chain. Non-local gates are executed by virtual SWAP chains that reorder the MPS sites, with each local SVD at cost $\mathcal{O}(\chi^3)$ and a gate of span $\ell$ requiring $\mathcal{O}(\ell)$ such SVDs. On $L=4$ the worst-case \textsc{ccx} span is $\sim\!15$ sites, set by the column-interleaved ordering combined with the torus wrap-around between periodic-boundary plaquettes [\Cref{fig:model}(a)]. After each correction round the MPS norm is restored numerically to absorb truncation drift. In both production runs ($\chi_{\max}=96$ for \Cref{sec:multiround} and $\chi_{\max}=256$ for \Cref{sec:threshold}), the observed bond dimension remained strictly below the cap on every trajectory, so the reported fidelities are not truncation-limited.

The code's ground state is prepared with a deterministic Clifford circuit~\cite{gottesman1997stabilizer,nielsen2010quantum}. Because the toric code is a CSS code~\cite{calderbank1996good,steane1996error}, the product state $\ket{0}^{\otimes n_d}$ is already a $+1$ eigenstate of every $Z$-type stabilizer $B_p$; it remains to project it into the $+1$ eigenspace of the $X$-type stabilizers $A_s$. We build the $(n_a/2)\!\times\!n_d$ binary check matrix $H_X$ over $\mathrm{GF}(2)$, reduce it to row-echelon form by Gaussian elimination~\cite{cleve1997efficient} (giving rank $n_d/2-1=7$ for $L=4$ after accounting for the global constraint $\prod_s A_s=I$), and convert each echelon row with pivot at column $j$ into a Hadamard on qubit $j$ followed by CNOTs from $j$ to the rest of the row's support. The resulting Clifford circuit (7 Hadamards, 35 CNOTs) takes $\ket{0}^{\otimes 16}$ into the simultaneous $+1$ eigenstate of every stabilizer, with numerical verification $\avg{A_s}=\avg{B_p}=+1$ and logical state $\ket{00_L}$.

\subsubsection{Noise model and Monte-Carlo unraveling}
\label{app:noise_full}

The code-level noise model used for the data of \Cref{sec:multiround,sec:threshold} consists of two ingredients: an i.i.d.\ depolarizing channel on each data qubit and a projective optical-pumping reset on the active ancilla species after each sub-cycle. The depolarizing channel is given by \Cref{eq:depolarizing} of \Cref{sec:multiround}: with probability $p$ each data qubit undergoes an event that applies $X$, $Y$, or $Z$ with probability $p/3$ each, and is untouched with probability $1-p$. Each correction round comprises sequential $X$- and $Z$-error sub-cycles with an ancilla reset after each sub-cycle (\Cref{sec:syndrome_design}; \Cref{fig:model}(b)); the depolarizing channel fires once at the start of the round, before the sub-cycles run. Gates are taken to be ideal so as to isolate the protocol's response to unstructured single-qubit errors; the gate-level error structure quantified in Methods~\ref{app:pulse_single} is treated as part of the same per-round budget $p$.

We model the optical-pumping reset of \Cref{sec:gate_design} as the dissipative CPTP channel (amplitude damping with $\gamma=1$)
\begin{equation}
    \begin{aligned}
        &\mathcal{R}(\rho) = K_0\,\rho\,K_0^{\dagger} + K_1\,\rho\,K_1^{\dagger}, \\
        &K_0 = \ket{0}\!\bra{0}, \qquad K_1 = \ket{0}\!\bra{1},
    \end{aligned}
    \label{eq:reset_channel}
\end{equation}
which deterministically maps any ancilla state to $\ket{0}$ and removes its coherence with the data. Because the simulator stores a pure MPS, $\mathcal{R}$ is realized via the standard quantum-jump (Monte-Carlo-wavefunction) unraveling of open-system dynamics~\cite{dalibard1992wave,molmer1993monte,carmichael1993open}: the Kraus index is sampled at Born weight $p_a=\|K_a\ket{\psi}\|^2$ and the post-jump state $K_a\ket{\psi}/\sqrt{p_a}$ is retained on every trajectory. One \emph{Monte-Carlo trial} (\Cref{sec:multiround}) is one such trajectory: initialised in $\ket{00_L}$, propagated through $n_{\rm rounds}$ noisy correction cycles, and terminated at a final pure MPS $\ket{\psi_{\rm corr}}$. Logical observables are estimated as sample averages over independent trials, with $95\%$ Wilson-score confidence intervals~\cite{wilson1927probable}.

%%%%%%%%%%%%%%%%%%%%%%%%%%%%%%%%%%%%%%%%%%%%%%%%%%%%%%%%%%%%%%%%%%%%%%%%%%%%%%%
% Back matter required by npj Quantum Information
%%%%%%%%%%%%%%%%%%%%%%%%%%%%%%%%%%%%%%%%%%%%%%%%%%%%%%%%%%%%%%%%%%%%%%%%%%%%%%%

\section*{Data availability}
The numerical data underlying every figure in this work are bundled together with the source
code in a single Zenodo archive at
\url{https://doi.org/10.5281/zenodo.20301590}.
The data files are released under the Creative Commons Attribution 4.0
International (CC BY 4.0) licence. Any additional data are available
from the corresponding author on reasonable request.

\section*{Code availability}
All simulation code supporting the
findings of this study is openly available at
\url{https://github.com/wanda0929/SelfCorrectingRydberg-code} and is
archived together with the underlying data on Zenodo at the same DOI
(\url{https://doi.org/10.5281/zenodo.20301590}), corresponding to the
tagged release \texttt{v1.1}. The repository is released under the MIT licence
and bundles three independently installable sub-packages:
(i) \textsc{triqg} (Python), the gate-level Lindblad and Choi-matrix
fidelity simulator used in Methods~\ref{app:pulse_single} to produce
\Cref{fig:gate_fid_bar};
(ii) \textsc{omqutensor} (Python), the matrix-product-state code-level
simulator of the rotated toric code used in Methods~\ref{app:simulation} to
produce \Cref{fig:ec_results}; and
(iii) \textsc{PulseODE} (Julia), the nine-atom Schr\"odinger-equation
integrator used in Methods~\ref{app:pulse_verify} to produce
\Cref{fig:pulse_sim}.
The upstream simulation packages remain under
active development at their primary repositories; the contents of the
\texttt{v1.1} archive correspond to the version used in this work.

% Acknowledgements: original \begin{acknowledgments} block preserved verbatim.
% The env shim in the preamble emits a \section*{Acknowledgements} header.
\begin{acknowledgments}
We thank Prof. Guoyi Zhu for insightful discussions and guidance during the
early stages of developing this protocol. We also thank Xiangyu Chen, Yang Qian, and Zhongyi Ni for helpful project discussions. This work was partially supported by the National Key R\&D Program of China (Grant No. 2024YFB4504004), the National Natural Science Foundation of China under grant nos. 12404568, the Quantum Science and Technology-National Science and Technology Major Project (Grants No. 2021ZD0301703), and the Shenzhen Science and Technology Program (KQTD20200820113010023).

\end{acknowledgments}

\section*{Author contributions}
H.W.\ designed the protocol and the pulse sequences, performed the
pulse-level Lindblad and MPS code-level simulations, and drafted the
manuscript. Y.Z.\ contributed to the simulation codebase. X.D.\ advised
on the pulse design and pulse-level simulations. J.L.\ supervised the
study and advised on the tensor-network simulations. All authors
discussed the results and contributed to the manuscript.

\section*{Competing interests}
The authors declare no competing financial or non-financial interests.

\end{document}